# Tuning phase-stability and short-range order through Al-doping in $(CoCrFeMn)_{100-x}Al_x$ high entropy alloys

Prashant Singh[1], Amalraj Marshal[2], Andrei V. Smirnov[1], Aayush Sharma[3], Ganesh Balasubramanian[4], K. G. Pradeep[2] and, Duane D. Johnson[1,5]

[1]Ames Laboratory, United States Department of Energy, Ames, IA 50011, USA

[2]Materials Chemistry, RWTH Aachen University, 52074 Aachen, Germany

[3]Department of Mechanical Engineering, Iowa State University, Ames, IA 50011, USA

[4]Dept. of Mechanical Engineering & Mechanics, Lehigh University, Bethlehem, PA 18015, USA

[5]Department of Materials Science & Engineering, Iowa State University, Ames, IA 50011, USA

**Abstract:**

**For $(CoCrFeMn)_{100-x}Al_x$ high-entropy alloys, we investigate the phase evolution with increasing Al-content ($0 \leq x \leq 20$ at.%). From first-principles theory, the Al-doping drives the alloy structurally from *FCC* to *BCC* separated by a narrow two-phase region (*FCC+BCC*), which is well supported by our experiments. We highlight the effect of Al-doping on the formation enthalpy and electronic structure of $(CoCrFeMn)_{100-x}Al_x$ alloys. As chemical short-range order (SRO) in multicomponent alloys indicates the nascent local order (and entropy changes), as well as expected low-temperature ordering behavior, we use thermodynamic linear-response within density-functional theory to predict SRO and ordering transformation and temperatures inherent in $(CoCrFeMn)_{100-x}Al_x$. The predictions agree with our present experimental findings, and other reported ones.**

**Introduction**

High-entropy alloys (HEAs) [1–3], which are based on the concept of high configurational entropy opposing the formation enthalpy of the alloy, continue to draw interest from the simple structures and novel properties that can appear [1,2]. As yet, several HEAs have been reported with body-centered cubic (BCC), face-centered cubic (FCC), and hexagonal closed-packed (HCP) [4,5,6] structures. With simple ground-state crystal structures, the HEAs offer a new class of materials with technologically promising properties potentially tuned using new alloy design strategies. Properties include high hardness [7], outstanding wear resistance [8], good fatigue lifetimes [9], excellent high-temperature strength [10], good thermal stability [11], and good oxidation resistance [12]. With superior radiation resistance, transition-metal-based HEAs, as compared to conventional single-phase Fe-Cr-Ni austenitic stainless steels, have attracted significant attention as high-temperature materials candidates for nuclear-energy applications [13].

Chemical complexity is a key feature of concentrated HEAs, particularly near equiatomic concentrations. Several studies have shown that additions of certain elements into transition-metal (TM) based HEAs may have strong effect on the microstructure and mechanical properties [14]. Adding 'Al' for example, the as-cast structure tends to evolve from single *FCC* phase to a mixture of FCC+BCC (duplex) phase to a single *BCC* phase [15,16]. In particular, the effect of Al content on the structural and tensile properties of CrMnFeCoNi system was carefully scrutinized [17], indicating that there is a structural phase transition from *FCC* to *BCC* with increasing Al content. Similar structural transitions in $Al_x$CrCoFeNi were found in experimental [15,16] and theoretical studies [18,19]. Unlike conventional Fe-based alloys, alloying becomes more complex and pivotal in designing novel HEAs with desired properties. For example, phase-stability and short-ranged order (SRO) in such alloys is very crucial, which has been shown to arise from the chemical complexity from competing pair correlations [18].

Here, we provide a systematic study of phase formation, structural, and magnetic stability of $(CoCrFeMn)_{100-x}Al_x$ HEAs from perspectives of varying Al-concentration. For selected set of alloys, i.e., x=0, 5 and 20 at.% Al, we perform first-principles KKR-CPA calculations to predict global (formation enthalpies) and local stability (short-range ordering (SRO)). In addition, we validate theoretical predictions with experimental results obtained for these HEAs.

**Results and Discussion**

*Phase formation and Structural Stability:* We predict the phase stability of $(CoCrFeMn)_{100-x}Al_x$ HEAs from the total-energy calculations of potential ground-state structures and finding the minimum via the formation enthalpies. Previous theoretical and experimental observations show that small Al-doping to TM-HEAs promotes the formation of stable *FCC/BCC*/duplex phases [3,18]. The Al-addition is also found to increase the equilibrium volume of the solid solution, which is consistent with the fact that atomic-radii of Al is larger than those of the other elemental components.

We address phase stability in Fig. 1(a) for *FCC/BCC*-phase of $(CoCrFeMn)_{100-x}Al_x$ alloy at 0 K. We find that the *FCC* phase is stable at x=0 at.% (no Al), a two-phase region (*FCC+BCC*) is lowest in enthalpy for $0 < x \leq 10$ at.% ("Al-poor" region), and a pure *BCC* phase is found for x > 10 at. % ("Al-rich" region). In Fig. 1(a), it is clear that the structural energy difference between ideal *FCC and BCC* lattices vanishes at x ≈ 4%Al. However, the common tangent to the formation enthalpy curves establishes that there is a two-phase (*FCC+BCC*) region (x < 10 at.%) that lowers the overall enthalpy of the homogenous system into a weighted mixture of two phases, where (*FCC+BCC*) equilibria may occur when the chemical potentials of the two phases become equal). The theoretical predictions are compared to the experimental results of (CoCrFeMn)$_{100-x}$Al$_x$ HEAs at various (0, 5, 20 at. %) Al concentrations, as shown in Fig. 1(b-d) XRD pattern. We observe predominant peaks in Fig. 1(b) for 0% Al corresponding to *FCC*. A tetragonal phase (Cr-based, as per ICDD JCP2 database PDF 09-0052, 05-0708) was also identified. In Fig. 1(c) for 5 at.%Al, similar to CoCrFeMn, multiple phases including *FCC* and a tetragonal phase (Cr based, as per ICDD JCP2 database) were observed along with a secondary *BCC*-phase. However, the XRD pattern of the equiatomic CoCrFeMnAl (i.e., 20 at.% Al) in Fig. 1(d) clearly shows single BCC phase. The role of Al in stabilizing the *BCC* phase was similar to the many reports for Al-containing HEA, where *BCC* phase is stabilized with increasing Al concentration [20-26].

Focusing on (CoCrFeMn)$_{100-x}$Al$_x$ at x=0 at.% (no Al) with single *FCC* phase, at x=5 at.% (Al-poor region) exhibting two-phase (*FCC+BCC*), and at x=20 at.% (Al-rich region) with single *BCC* phase, we calculate each alloy's electronic structure and short-range order, and compared with our experimental results. The calculated lattice constants for (CoCrFeMn)$_{100-x}$Al$_x$ are 3.54 Å (*FCC*) at 0 at.%Al, 3.55 Å (*FCC*) and 2.83 Å (*BCC*) at 5 at.%Al and 2.84 Å (*BCC*) at 20 at.%Al, which agree well with our experimental findings of 3.59 Å (0 at.% Al, *FCC*), 3.58 Å (5 at.% Al, *FCC*) and 2.86 Å (5 at.% Al, *BCC*), and 2.86 Å (20 at.% Al, *BCC*), respectively. Although changing Al concentration shows very little effect on lattice constants (either calculated or experiment), the calculated bulk modulus reduces significantly from 247.1 GPa at 0 at.%Al, to 218.3 (*FCC*)/209.1 (*BCC*) at 5%Al, and to 169.5 GPa at 20%Al. This happens because Al additions reduces the alloy's interstitial electron number (electronic density) [27], which controls the moduli. The calculations at 0 K are helpful to understand, characterize and predict mechanical properties [18] and enables selection of materials for further studies.

*Magnetic stability:* For (CoCrFeMn)$_{100-x}$Al$_x$ HEAs, we study non-magnetic (NM), paramagnetic (PM), and ferro-magnetic (FM) states. The PM phase is addressed within the disorder-local moment (DLM) approximation [20]. For 0% Al, *FCC* CoCrFeMn has a NM ground state that is energetically more stable than DLM and FM. Adding %Al in CoCrFeMn HEA becomes magnetic and DLM or FM phase is lower in energy than NM phase. For 5%Al, DLM phase is energetically more stable than NM or FM phase. In DLM phase, Fe/Mn/Co possess non-zero site moments. Increasing from 0 to 5 at.%Al (see Fig. 1(a)), the alloy shows a mixed (*FCC+BCC*) phase, so we analyze magnetic behavior of both *FCC* and *BCC* phases. In FCC phase, only Fe possesses small DLM moment of 0.52 $\mu_B$, while in *BCC* phase Fe/Mn/Co shows DLM moment of 2.02/0.92/1.00 $\mu_B$. For 20 at.% Al (equiatomic case), the alloy forms single *BCC* phase, with DLM moment on Fe/Mn/Co reduced slightly to 1.91/0.98/0.79 $\mu_B$ relative to 5 at.%Al *BCC* phase, and average total local moment of 0.74 $\mu_B$ at 0 K, with net zero magnetization. The experimentally measured total magnetization of the equiatomic CoCrFeMnAl BCC-stabilized HEA is 0.58 μ$_B$ at 50 K (0.45 μ$_B$ at 300 K). With increasing at.%Al, (CoCrFeMn)$_{100-x}$Al$_x$ undergoes a transition from NM-*FCC* to DLM/FM-(*FCC+BCC)* to FM-*BCC* state. The at.%Al has a strong impact on the magnetic behavior of the HEA.

To assess the Curie temperature ($T_c$), we consider differences in the total energy/atom E$_{DLM}$ and E$_{FM}$ for DLM and FM states, respectively. In contrast to segregating systems [28], the transition temperature in ordering systems (with negative formation enthalpies) are dictated mainly by energy differences [29]. Thus, for magnetic ordering, mean-field theory considerations gives an estimated $T_c = \frac{2}{3}[E_{DLM} - E_{FM}]/k_B$, where $k_B$ is Boltzmann's constant. For equiatomic *BCC*-CoCrFeMnAl, the calculated $E_{DLM} - E_{FM}$ is 3.61 mRy, giving $T_c = 380\ K\ (107^oC)$. In Fig. 2 for BCC (CoCrFeMn)$_{100-x}$Al$_x$, we plot the estimated $T_c$ versus $x$ %Al (x=5-30%Al) [30], and include an upper bound (see Methods Section).

*Valence electron count (*VEC*):* The VEC can be a key parameter found self-consistently in first-principles electronic-structure calculations, which is obtained by integrating the density of states (DOS) of the valence states over occupied states (up to the Fermi energy, $E_F$). VEC is correlated directly to the stability of *FCC* or *BCC* phases. To best of our knowledge, in solid solution forming HEAs, there are no exceptions to the trend that a higher VEC favors the *FCC* phase and a lower VEC favors the *BCC* phase. This can be justified as follows: *BCC* forms for 4<VEC<6 [3, 31], as stability increases when bonding *d*-states fill and maximal when half-filled (VEC~6); anti-bonding states fill roughly when VEC>6 and stability decreases. For 6.8<VEC<8 other phases compete (e.g., σ phase), and *FCC* become stable for VEC>8 [3, 31]. The threshold VEC values of 6.87 and 8.0 still seem to be a reasonable guide to stabilize *FCC* or *BCC* HEAs. In the (CoCrFeMn)$_{100-x}$Al$_x$, *FCC* HEA is more stable than *BCC* at a VEC~7.5 (for x= 0 at.%), although the formation enthalpy is positive (not favorable except at higher temperatures due to entropy), while *BCC* begins to be stabilized (i.e., negative formation enthalpy) at VEC=6.6 (x=20 at. %), as seen in Fig. 1(a). Hence, an *FCC+BCC* phase exists at an intermediate VEC (e.g., 7.275 at x=5 at.%), although the formation enthalpy is still positive. Thus, we see roughly the same VEC limit for (CoCrFeMn)$_{100-x}$Al$_x$ as the empirically defined solid-solution phase limit, with exceptions already noted for Mn-containing HEAs [3]. For (CoCrFeMn)$_{100-x}$Al$_x$, empirical and first-principles analysis correlate well in predicting relative phase stability, which is not always the case, as in refractory alloys [31].

To shed more light on the effect of Al-doping on electronic structure in an alloy (Fig. 3), we show configurationally averaged DOS and BSF, i.e., the electronic dispersion, for *FCC* and *BCC* (CoCrFeMn)$_{100-x}$Al$_x$ for 0, 5, and 20 at.% Al. For *FCC* CoCrFeMn, the DOS and BSF both exhibit sharp structure at lower (filled) energies, similar to that of a pure metal. For 5 and 20 at.%Al, the alloys show disorder broadening both near and far from the Fermi energy (E$_F$) due to increased scattering from Al alloying and DLM. The presence of finite local moment with increased Al concentration provides both chemical and magnetic disorder for enhanced scattering. This effect is seen in the DOS, which shows less structure with alloying plus DLM due to disorder broadening. Clearly, in Fig. 3, the deep lying states move with increased %Al, which shows hybridization between Al *s*-states and Fe/Co/Cr bonding 3*d*-states. The shift in dispersion well below E$_F$ (easily seen in the low-energy *s*-states) also indicates enhanced stability of the alloy due to lowering states in energy from hybridization with Al. In (CoCrFeMn)$_{100-x}$Al$_x$, Al works as an *BCC* stabilizer by hybridization/band-filling that enhances disorder and lowers bonding states [18]. For equiatomic HEA (20 at.%Al), see Fig. 1(a), the negative formation energy indicates the favorability for elements to mix, in contrast to the alloys with lower Al concentration (0 and 5 at.% Al), which shows segregation.

*Short-range order:* For *FCC*-CoCrFeMn (0 at.%Al), in Fig. 4(I), we plot the chemical interchange energies [$S^{(2)}_{\alpha\beta}(\mathbf{k})$] and Warren-Cowley SRO parameters [$\alpha_{\alpha\beta}(\mathbf{k})$] at 1.15 of the calculated spinodal temperature [T$_{sp}$=315 K (42$^0$C)]. $S^{(2)}_{\alpha\beta}(\mathbf{k})$ shows a very weak Co-Cr peak at $\mathbf{k}_0$=W={1,½,0} in the *FCC* Brillouin zone. The peak in SRO at W shows a weak ordering, which indicates the possibility of very low-temperature (below T$_{sp}$=315 K) tetragonal phase (DO$_{22}$), which also observed in as secondary phase in XRD measurements (see Fig.1(b)). However, the peak in $\alpha_{\alpha\beta}(\Gamma)$ has a dominant segregation for Fe-Co pairs, followed very closely by Mn-Cr pairs. Furthermore, Fe-Mn, Fe-Cr, and Mn-Co pairs show flat weight, indicating tendency of equal probability of mixing. The Co-Cr pair, which drives SRO, indicates the tendency of Cr to enrich over Co in the low-T tetragonal phase. Experiments, in Fig. 4(II), also indicate phase separation and the presence of tetragonal phase.

In Fig. 4 (II) (a), we show the backscattered electron (BSE) image of the 0 at.%Al alloy representing the phase-separated microstructure. The presence of two-phase regions (inferred from the phase contrast) is consistent with the XRD results in Fig. 1(b). EDX mapping of the corresponding area, in Fig. 4II (a), reveal regions (1 and 2) with different chemical concentration, where region 1 has larger Cr concentration than the region 2. The quantification of these two chemically distinct regions is given in inset of Fig. 4. The Cr-rich region (1) has ~33 at.% Cr, with constant Mn and Fe concentrations (~24 at.% each), and was found to be depleted in Co. In contrast, region 2 (with ~23 at.% Cr) was (relatively) rich in Mn, Fe and Co. Similar separation of Cr-based phase has been widely reported in other Cr-containing transition-metal-based HEAs [17, 21,32]. The observed phase separation phenomenon is in good agreement with calculated SRO (see Fig. 4(I)). The thermal stability analysis of the multiphase CoCrFeMn HEA is shown in Fig. 5(I). An endothermic peak (at 1130°C), potentially corresponding to the decomposition of the observed Cr-rich phase (spinodal decomposition), was observed in addition to the solid-liquid phase transition (peak at 1317°C). Our mean-field estimate of the miscibility gap $T_{MG} = \Delta E_f/\Delta S_{conf}$ ~10.3 mRy/$k_B$ln4 ~1174 K (901°C) is similar to DSC measurements (1130°C).

Notably, however, the SRO-predicted $T_{sp}$ of 315 K (42°C) for *FCC* CoCrFeMn deviates from that found directly from the enthalpy estimate 1174 K (901°C) (where they should agree). We have tracked this discrepancy to variations of the ASA-only free energy used to estimate the SRO response functions (which will be corrected in the future by including contributions from interstitial Coulomb energy with cell shape and periodicity). The discrepancy mostly results in a decrease of the spinodal temperature in the *FCC* phase, and fortunately has less effect on the *BCC* phase. In short, the competing phase separating (clustering) (Γ) and weak-ordering (W) seen in SRO in Fig. 4(I) shows qualitative agreement with experiments.

For BCC (CoCrFeMn)$_{95}$Al$_5$, in Fig. 6(I), $S_{\alpha\beta}^{(2)}(\mathbf{k})$ and $\alpha_{\alpha\beta}(\mathbf{k})$ both have peak at $\mathbf{k}_0$=Γ=[000] in the BCC Brillouin zone. The peak at Γ indicates an infinite wavelength concentration wave indicative of phase separation. The $\alpha_{\alpha\beta}(\Gamma)$ has a dominant SRO peak in Cr-Al (followed by Co-Cr and Fe-Cr); however, the phase separation is actually driven by the Co-Al pair in $S_{\alpha\beta}^{(2)}$. Fig. 6(II) (a) shows the multiphase microstructure of the 5%Al HEA, which consists of a lamellar grain morphology. The corresponding elemental mapping by EDX of the same region reveals Cr-rich and Cr-depleted regions in Fig. 6(II)(b), similar to the CoCrFeMn in Fig. 5(II)(b). The Cr-rich region had ~29 at.% Cr, with constant Mn, Fe and Co concentration of ~22 at.%. The Cr-depleted region had ~20 at.% Cr with ~25 at.% of Mn, Fe and Co, which is slightly higher than the Cr-rich region.Whereas, the maximum Al was partitioned to the Cr-rich region.. The observed instability to phase separation at the 5 at.%Al is in good agreement with the theoretical predictions. The thermal stability evaluation of the phase separated 5 at.%Al HEA is shown in Fig. 6(II). The measured DSC curve shows a prominent endothermic peak along with the melting peak, similar to one observed for CoCrFeMn HEA in Fig. 5(I). The phase transformation could be due to the phase decomposition into the observed Cr-rich and Cr-poor regions. In contrast to 0 & 5%Al cases, equiatomic CoCrFeMnAl shows dominent peak in $S_{\alpha\beta}^{(2)}(\mathbf{k})$ and $\alpha_{\alpha\beta}(\mathbf{k})$ at $\mathbf{k}_0$=H=[001] as shown in Fig. 7(I) for the *BCC* Brillouin zone. Clearly, the Mn-Al pair drives the instability as shown in $S_{\alpha\beta}^{(2)}(\mathbf{k})$, however, the SRO has dominant peak in Mn-Cr. This apparent oddity (but correct) in multicomponent alloys arises from the probability sum rule (or optical theorem for particle conservation) [33].

The microstructure of the equiatomic HEA is displayed in Fig. 7(II) (a). Equiaxed grains with an average grain size of ~10 μm were observed. The EBSD phase mapping in Fig. 7(II)(b) of the corresponding area indicates the presence of single *BCC* phase. This further affirms calculated phase stability predictions and XRD observations in Fig. 1 (a) & (d). No indication of phase (chemical) separation was observed in the EDX mapping (not shown here) as well. However, atomic-scale investigation by APT reveals nano-scale chemical separation in the analyzed volume of 145 x 147 x 212 nm$^3$ (as observed in Fig. 8(a)), where Co- and Cr-rich regions were seen. Corresponding one-dimensional concentration profile taken along a cylindrical region of interest, of 10 x 10 x 180 nm$^3$ along the tip length shows anti-correlated fluctuations between Cr-Mn-Fe and Co-Al elements (Fig. 8(b)). Even though anti-correlated fluctuations are characteristics of spinodal decomposition [22], DSC curves reported for equiatomic CoCrFeMnAl in Fig. 5(III) had no signs of phase transition until the melting point. Figure 9 (inset) shows iso-concentration surfaces of Co and Cr, an intertwined morphology was observed between the Cr- and Co-enriched regions. Proxigram extending from Co-rich to Cr-rich region across the interface was plotted (Fig. 9) to determine the actual chemical composition of the separating phases. The Cr-rich region had a composition of $Co_7Cr_{31}Fe_{25}Mn_{25}Al_{12}$, while the concentration of the Co-rich region was $Co_{38}Cr_4Fe_{25}Mn_{25}Al_{33}$.

The SRO in Fig. 7(I) shows peak at H [001] with Cr-Mn as dominant pair in CoCrFeMnAl. The peak at H-point indicates the possibility of B2-type ordering. While no *B2* ordering was observed in our XRD data (Fig. 1(d)), a recent work on an equiatomic single crystal reports *B2*-type ordering [23]. The synthesis of a single crystal involved extremely low cooling rates, and resulted in the formation of nearly equilibrium phase; the Cr-rich region was disordered *BCC* and the Co-Al-rich precipitates were ordered *B2*, embedded in the disordered *BCC* matrix [23]. The two-phase structure is thus similar to that found in equiatomic AlCoCrFeNi HEA [22]. The *BCC/B2* coherent morphology is closely related to the lattice misfit between these two phases, which is induced by the spinodal decomposition at higher Al concentrations. The calculated SRO, which is based on the concept of phase separation, arises from the spinodal decomposition driven by the strongest SRO pair Mn-Cr, in Fig. 7(I), which leads to phase decomposition at $T_{sp}$. The Mn-Cr rich phase observed by experiments, further affirm our SRO predictions. For the $(CoCrFeMn)_{100-x}Al_x$ HEA, summary of phase stability and SRO predictions from calculations and experiments are listed in the Table. I.

From first-principles alloy theory, we have predicted the phase stability (formation enthalpy), magnetic stability and short-range order (SRO) properties of $(CoCrFeMn)_{100-x}Al_x$ HEAs (x ≤ 20%). The theoretical results agree well with our current and other reported experimental data. The $(CoCrFeMn)_{100-x}Al_x$ shows Al-dependent phase stability as the HEA transforms from *FCC* (x= 0 at.% Al) to *BCC* (x > 10 at.% Al) with two-phase *(FCC+BCC)* region in between. Al (*s-p* orbitals) hybridizing with the transition-metals (*d*-orbitals) plays a crucial role in structural stability and magnetic, elastic and chemical properties. The SRO calculations indicate the presence of complex phase at 0 at.%Al, phase separation at 5 at.%Al, and a partially-ordered B2 phase at 20 at.%Al, in agreement with the experimental data. KKR-CPA-based electronic-structure and linear-response SRO methods offers a quantitative theory-guided design strategy to tailor the structural, chemical, magnetic, and mechanical properties of novel multicomponent alloys by tuning the concentration of alloying elements.

**Methods**

**KKR-CPA calculations:** The KKR electronic-structure method is an all-electron Green's function method implemented within a scalar-relativistic approximation, i.e., spin-orbit is ignored beyond the core electrons. KKR is combined with the coherent-potential approximation (CPA) to address chemical disorder [34,35]. The screened-CPA is used to address Friedel screening from configurational charge-correlations [36]. Formally configurational averaging used in KKR-CPA require only 1-atom (2-atom) per cell for disorder alloys with *FCC*/BCC (HCP) structures for any arbitrary composition. Green's function integration used complex-energy contour on a 20-point Gauss-Legendre semicircular contour, taking advantage of analytic continuation to decrease dramatically solution times [37]. The generalized gradient approximation to exchange-correlation within density functional theory (DFT) was included through use of *libXC* libraries [38]. Formation energies ($E_f$), electronic density of states (DOS), and electronic dispersion are calculated within the atomic sphere approximation (ASA) with periodic boundary conditions (PBC) to incorporate interstitial electron contributions to Coulomb energy from all atomic Voronoi polyhedra. Brillouin zone (BZ) integrations for self-consistent charge iterations were performed with a 20×20×20 Monkhorst-Pack ***k***-point mesh [39], whereas a *50×50×50* mesh was used for the physical density of states (DOS) calculations. For Bloch spectral functions (BSFs), we used 300 ***k***-points along high-symmetry lines in the irreducible FCC/BCC BZ to visualize the electronic dispersion. As needed, we investigate HEAs with 3*d* magnetic elements in the paramagnetic state using the DLM approximation [20]. The DLM state, representing the state above the magnetic transition temperature (e.g., Curie temperature for ferromagnet), has finite local moments on an atomic site that is randomly oriented (4π steradian orientations), which give zero magnetization on average over orientations. Notably, the DLM state is distinct from a non-magnetic state (no magnetization per site) or from an ordered configuration with collinear moments on a site that sum to zero in a supercell, which is an antiferromagnetic or ferrimagnetic, rather than a paramagnet, state.

**Chemical SRO:** From KKR-CPA-based thermodynamic linear-response, we calculate the Warren-Cowley SRO parameters, $\alpha_{\mu\nu}(\mathbf{k};T)$, for μ-ν pairs [18]. Dominant SRO pairs are identified above the spinodal ($T_{sp}$) from the chemical stability matrix (pair interchange energy) $S^{(2)}_{\mu\nu}(\mathbf{k};T)$. At the $T_{sp}$, $\mathbf{\alpha}^{-1}_{\mu\nu}(\mathbf{k}_o; T_{sp})$ vanishes, signifying an absolute instability to chemical fluctuations and provides an estimate for the order-disorder temperature or miscibility gap [18,31,40-43]. $S^{(2)}_{\mu\nu}(\mathbf{k}; T)$ reveals the unstable (Fourier) modes with ordering wavevector $\mathbf{k}_o$, or phase separation (clustering) at $\mathbf{k}_o=(000)$. Currently, the SRO theory is coded only in a pure ASA formalism, so the KKR-CPA-ASA is used for all $S^{(2)}_{\mu\nu}(\mathbf{k}; T)$ calculations. Note: this ASA-only formalism neglects differences in interstitial Coulomb energy that vary with crystal structure (a potential source of discrepancy for SRO temperature scale).

**Curie Temperature Estimate**: In ordering systems, energy difference mainly determines the transition temperature [29]. Thus, to assess the Curie temperature (magnetic ordering), we consider a Heisenberg-like model in mean-field [30], and find that $T_c = \frac{2}{3}[E_{DLM} - E_{FM}]/k_B$, proportional to the energy difference between PM and FM states. Sometimes, as in a dilute magnetic semiconductor [30], it is appropriate to consider a non-magnetic element with concentration $c$ and $T_c = \frac{2}{3} \cdot \left[\frac{1}{1-c}\right] \cdot [E_{DLM} - E_{FM}]/k_B$, which serves here as an upper bound.

**Materials Preparation and Characterization:** CoCrFeMnAl based high entropy alloys (HEA) with three different Al concentrations (0, 5 and 20 at.%) were synthesized by arc melting of high purity (≥ 3N) metal powders of Co, Cr, Fe, Mn and Al, under an Ar atmosphere. To ensure compositional homogeneity, the alloys were re-melted five times prior to casting in copper moulds. The crystal structure of the three HEAs was determined by X-ray diffraction (XRD) using Cu-Kα radiation. A Siemens D5000 diffractometer operating in Bragg Brentano geometry at 40 kV was used to obtain the diffractograms. Alloy microstructure and the corresponding surface chemical distribution were mapped using a Zeiss SIGMA™ field emission scanning electron microscope (FE-SEM) equipped with an Oxford X-maxN™ energy dispersive X-ray spectroscopy (EDX) detector. An electron accelerating voltage of 20 kV was used, and the chemical quantification was performed considering only the Kα x-rays. Electron back scattered diffraction (EBSD) was performed to further verify phase formation and to visualize the microstructure along with their grain orientations. A Zeiss LEO 1530 FE-SEM with an EBSD detector was used for this purpose, and the analysis of EBSD data was performed with AZtecHKL software. The thermal stability of the HEAs were investigated using NETZSCH STA 449C differential scanning calorimeter (DSC). The heating and cooling curves were recorded at a reduced rate of 5 K/min until 1500°C in an Ar atmosphere. Local electrode atom probe (LEAP 4000X HR$^{TM}$) was used to study the three-dimensional elemental distribution on a near-atomic scale. The samples for atom probe tomography (APT) were prepared using a FEI Helios Nanolab 660 dual beam workstation. APT measurements were performed at the set temperature of 60 K in laser pulsing mode, with an applied pulse energy and frequency of 30 pJ and 250 kHz, respectively. APT data reconstruction and analysis were carried out using IVAS 3.6.10a software provided by Cameca Instruments.

**Acknowledgements**

Work supported by the U.S. Department of Energy (DOE), Office of Science, Basic Energy Sciences, Materials Science & Engineering Division. Research was performed at Iowa State University and Ames Laboratory, which is operated by ISU for the U.S. DOE under contract DE-AC02-07CH11358. Work by AS & GB supported by the Office of Naval Research under grant N00014-16-1-2548 and N00014-16-1-2484.

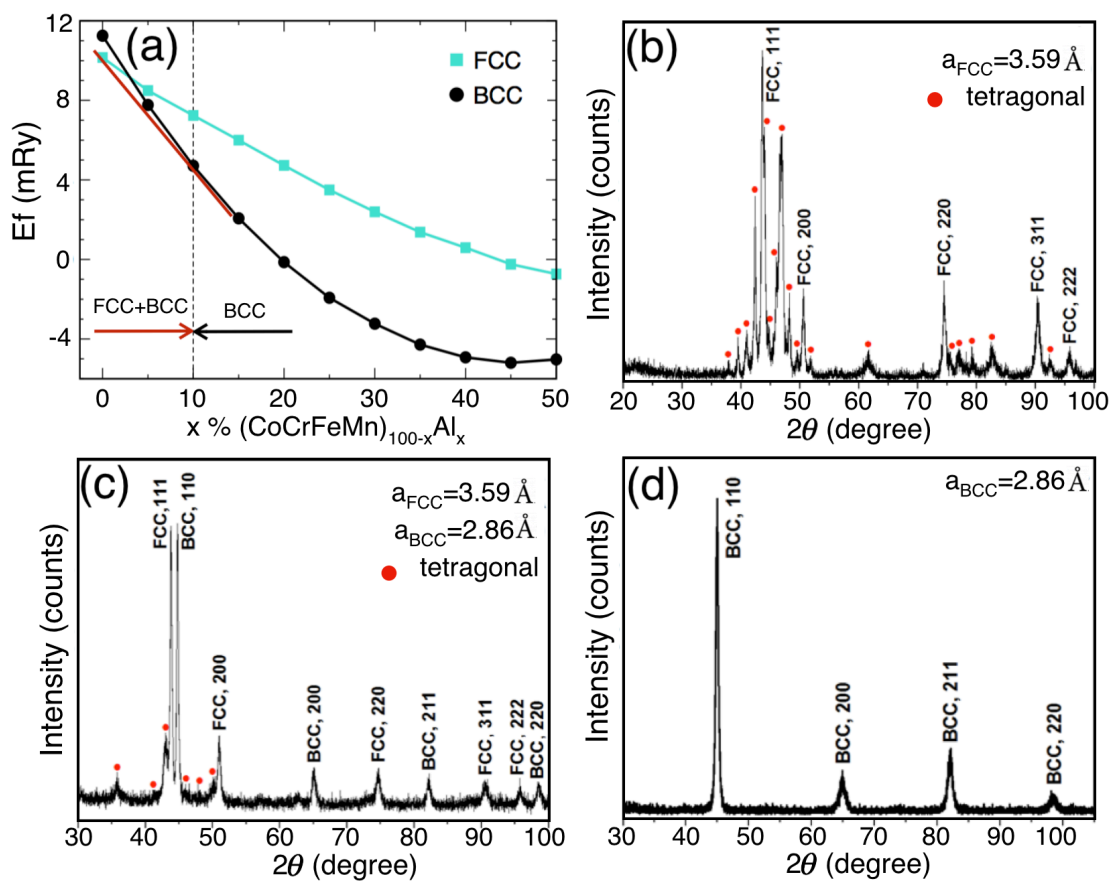

Fig. 1. (a) Formation energy (mRy/atom) versus at.% Al in *FCC* and BCC (CoCrFeMn)$_{100-x}$Al$_x$. Common tangent line (red line) highlights mixed (*FCC* + BCC) phase for x < 10 at. %Al. Adding more Al stabilizes the BCC phase. XRD pattern of the as-cast (CoCrFeMn)$_{100-x}$Al$_x$ HEA, for (b) x=0, (c) x=5, and (d) x=20 at.%Al.

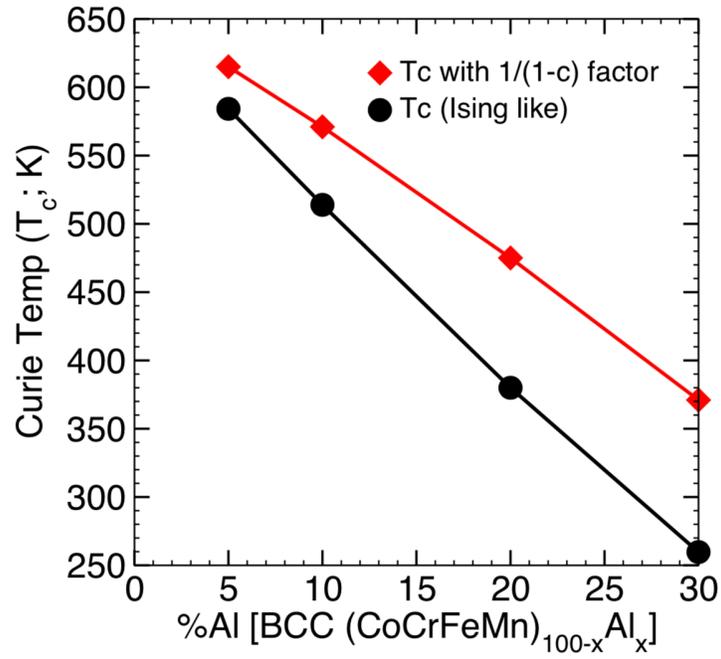

Fig. 2. Curie temperatures versus %Al for BCC $(CoCrFeMn)_{100-x}Al_x$ estimated from mean-field approximation [28-30], see Methods Sections.

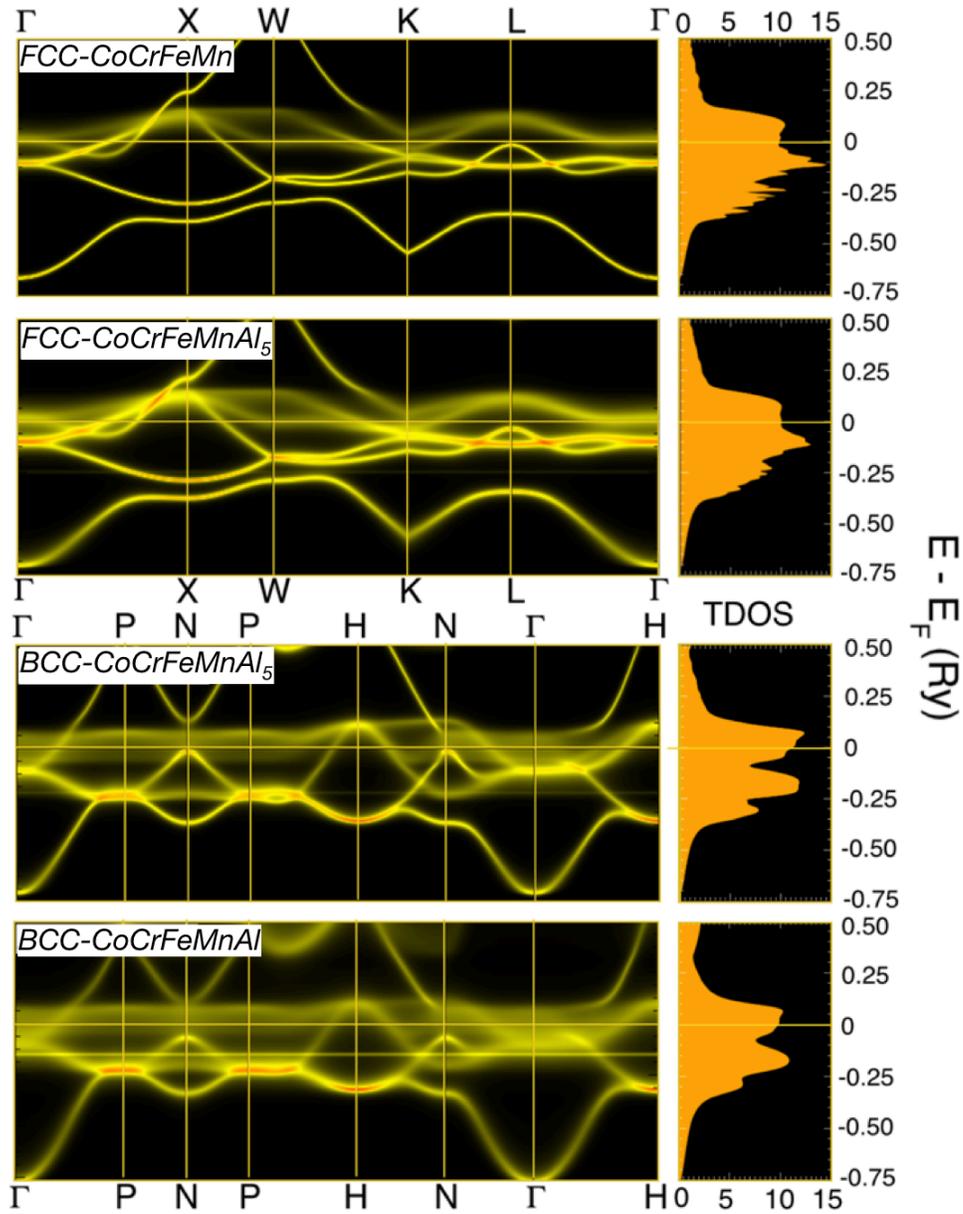

Fig. 3. Electronic dispersion (left) from the Bloch-spectral function (BSF) of (**CoCrFeMn**)$_{100-x}$Al$_x$ at x=0, 5, 20 at.% Al along high-symmetry directions of *FCC* and *BCC* Brilluoin zone and (right) total density of states [states/(Ry-atom-spin)].

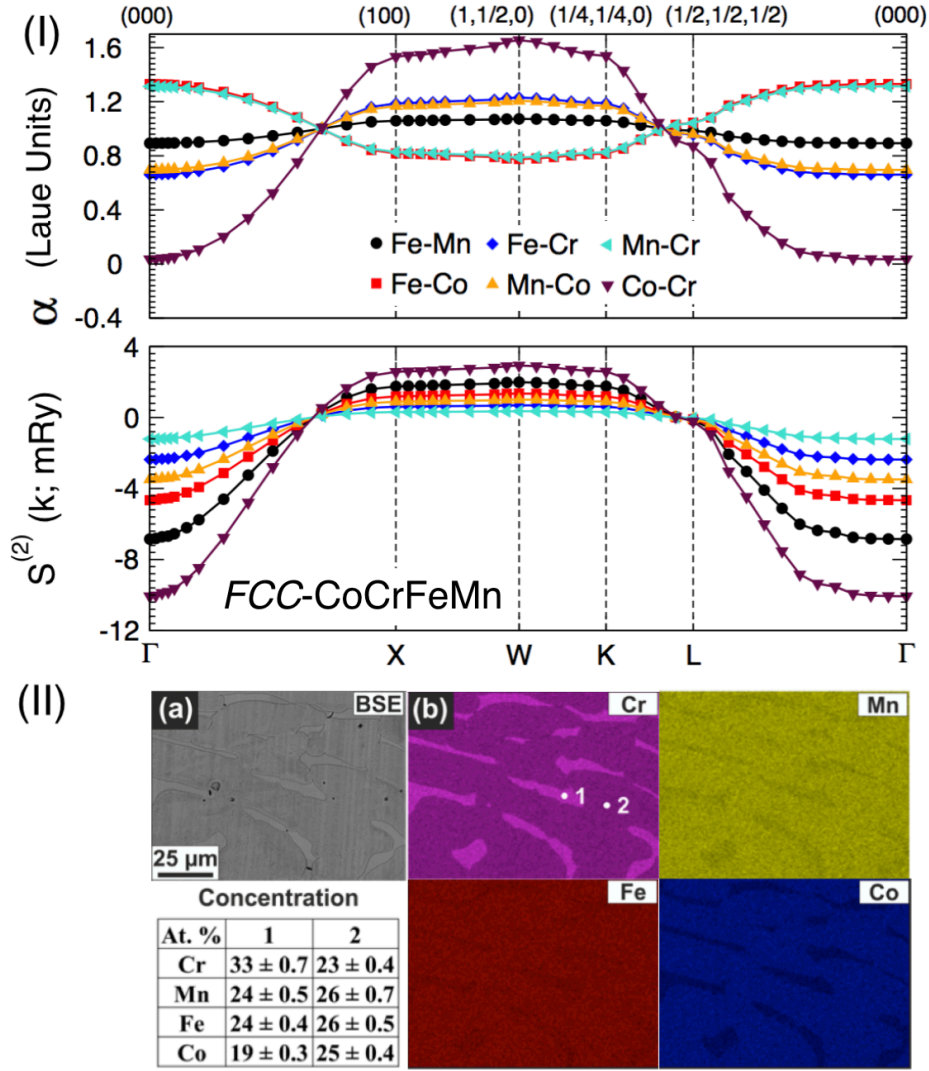

Fig. 4. (I) For equiatomic *FCC* CoCrFeMn, (top) Warren-Cowley parameters $\alpha_{\alpha\beta}$ (k; 1.15T$_{sp}$) with T$_{sp}$=315 K (42°C) and (bottom) pair interchange energies, $S^{(2)}_{\alpha\beta}$ (k; 1.15T$_{sp}$) plotted along high-symmetry directions of Brillouin zone (Γ-X-W-K-L-Γ). W-mode (1 ½ 0) ordering instability is driven by Co-Cr in $S^{(2)}_{Co-Cr}$(W), however, closely competing pairs (Fe-Co; Mn-Cr) peak in the observable $\alpha_{\alpha\beta}$(Γ). (II) (a) BSE microstructure and (b) corresponding EDX maps of CoCrFeMn. Composition of selected spots (1 and 2) in the microstructure is presented in the inset table.

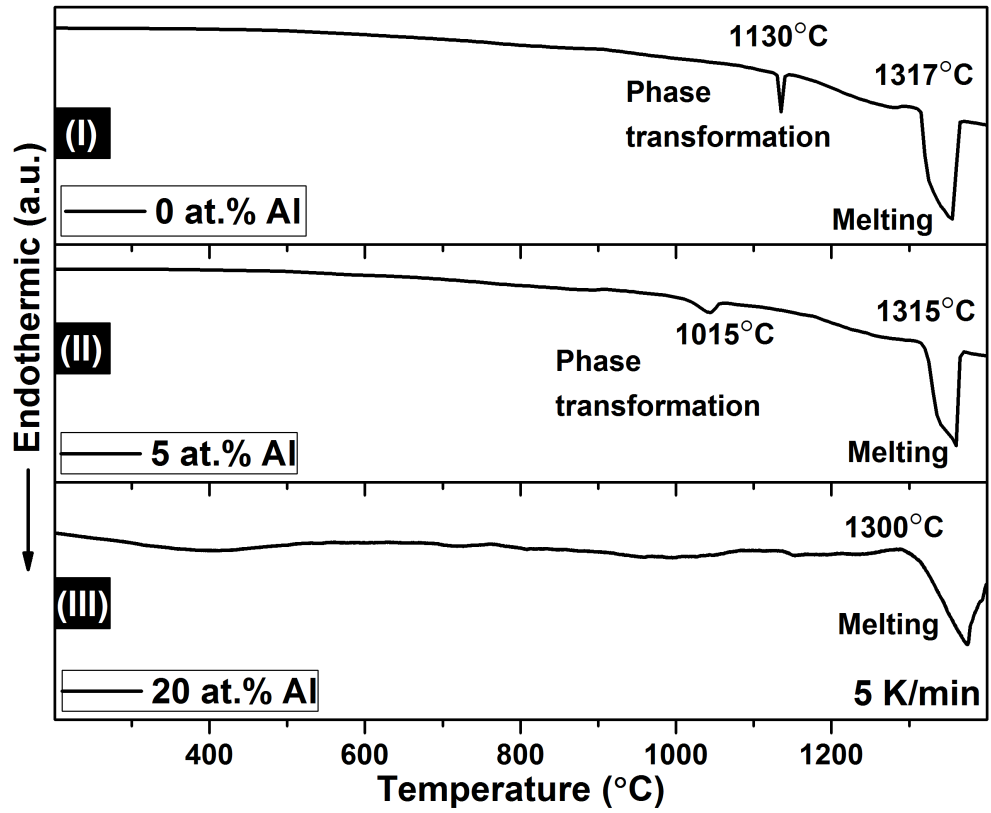

Fig. 5. DSC trace of (CoCrFeMn)$_{100-x}$Al$_x$ for x: (I) 0,(II) 5, and (III) 20 at.%Al (from Ref [35]).

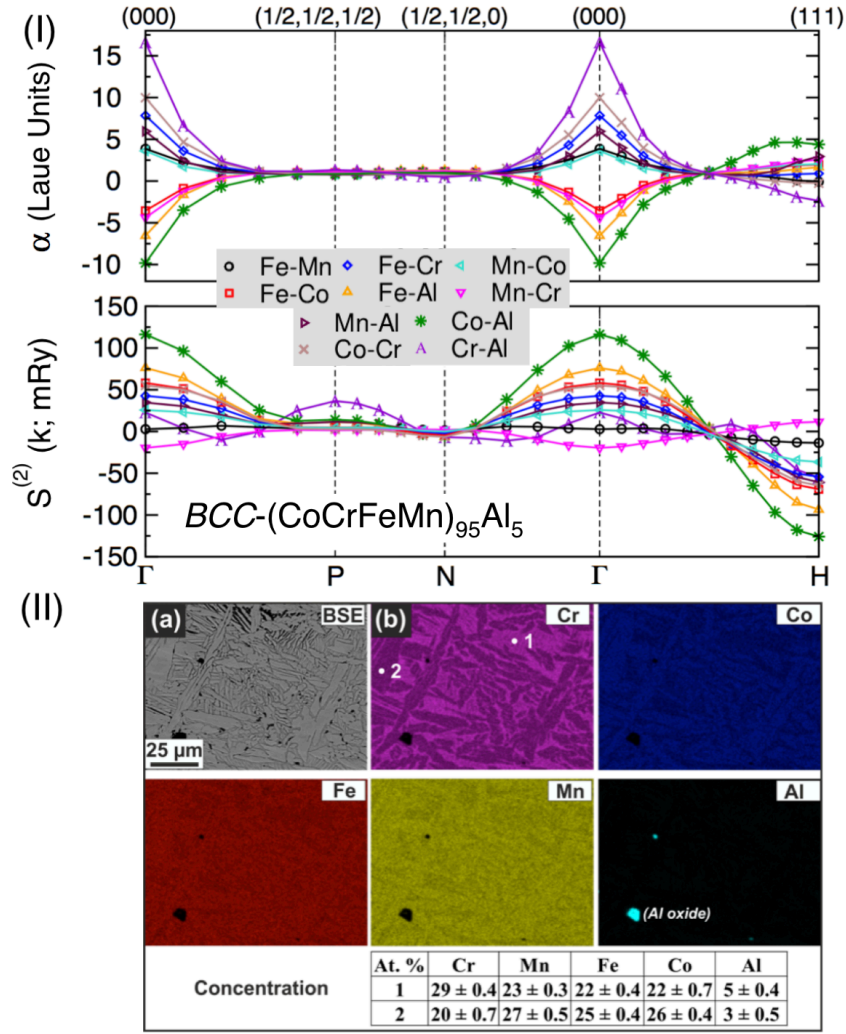

Fig. 6. (I) For *BCC* (**CoCrFeMn**)$_{95}$Al$_5$, (top) Warren-Cowley $\alpha_{\alpha\beta}$ (**k**; 1.15T$_{sp}$) with T$_{sp}$=1410 K (1137°C) and (bottom) pair interchange energies, S$^{(2)}_{\alpha\beta}$ (k; 1.15T$_{sp}$) along high-symmetry directions of Brillouin zone (Γ-P-N-Γ-H). Γ-mode (000) instability to clustering is driven by Cr-Al in S$^{(2)}_{Cr-Al}$(Γ), and manifested by Co-Al peak in $\alpha_{\alpha\beta}$(Γ). (II) (a) BSE microstructure and (b) corresponding EDX maps of the 5 at.%Al HEA. Composition of selected regions (1 and 2) in the microstructure is shown in the inset table. Dark spots in (a) corresponds to (Al) oxide inclusions.

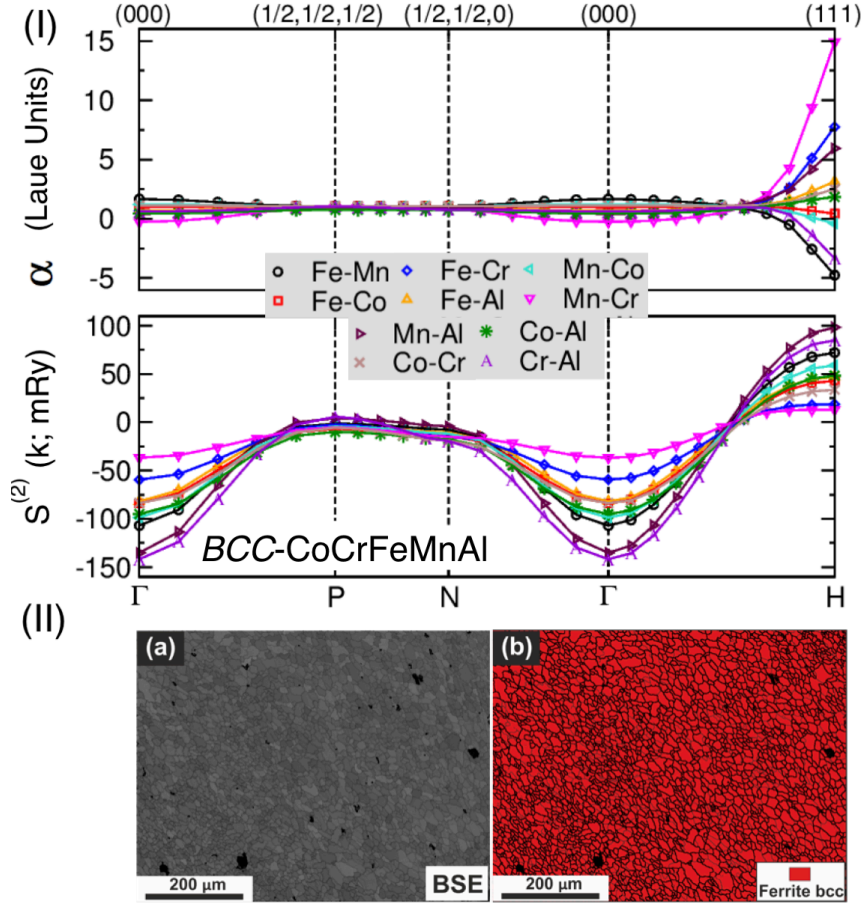

Fig. 7. (I) For equiatomic *BCC* CoCrFeMnAl HEA, (top) Warren-Cowley parameters $\alpha_{\alpha\beta}$ (k; 1.15$T_{sp}$) with $T_{sp}$=1360 K (1087°C), and (bottom) pair interchange energies, $S^{(2)}_{\alpha\beta}$ (k; 1.15$T_{sp}$) along high-symmetry directions of Brillouin zone (Γ-P-N-Γ-H). H-mode [001] ordering instability is driven by Mn-Al in $S^{(2)}_{Mn-Al}$(H), and manifested by Cr-Mn peak in $\alpha_{\alpha\beta}$ (H). (II) For CoCrFeMnAl, (a) BSE microstructure and (b) EBSD phase map.

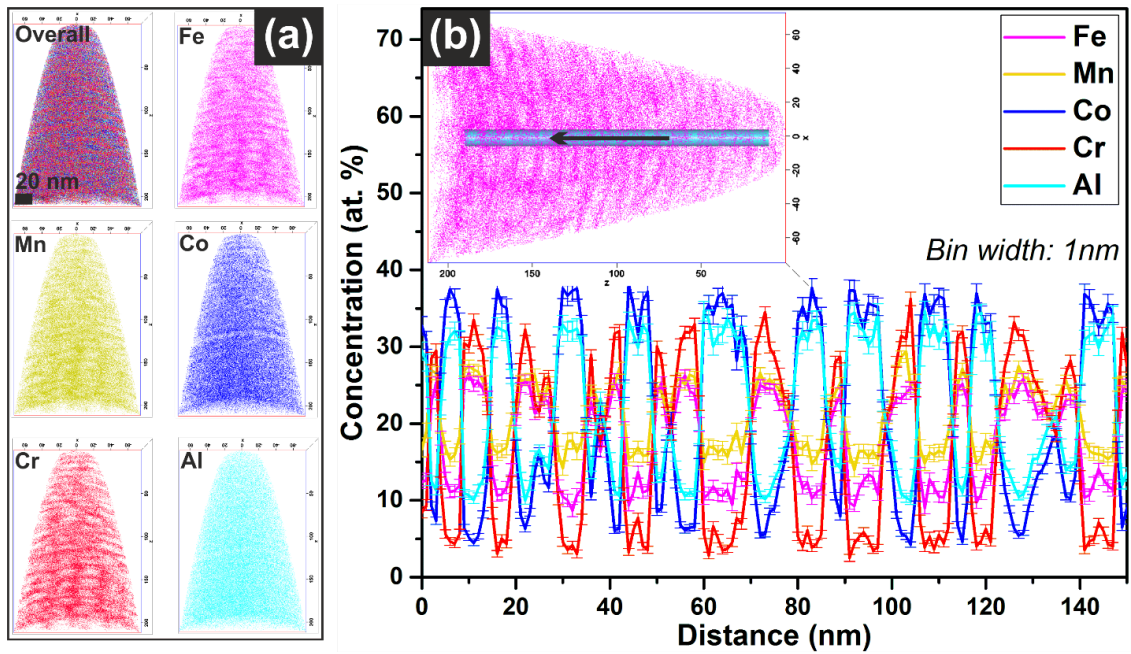

Fig. 8. For CoCrFeMnAl, APT analysis showing (a) three-dimensional reconstruction of Co, Cr, Fe, Mn and Al atom positions. (b) One-dimensional concentration profile of the alloying elements taken along a cylinder (shown in the inset) of 10 nm diameter with 1 nm bin width.

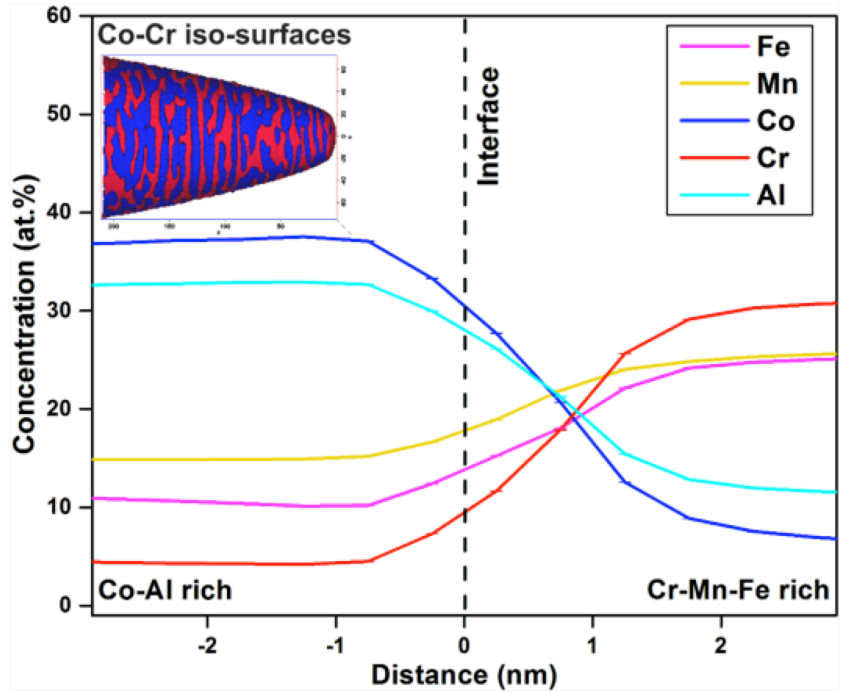

Fig. 9. APT analysis of CoCrFeMn Al showing isosurfaces highlighting the (30 at.%) Co and (20 at.%) Cr regions (inset), and the corresponding proximity histogram of 0.5 nm bin width across the separating phases.

| HEA → | 0 at.% Al | 5 at.% Al | 20 at.% Al |
|---|---|---|---|
| **Calculated phase stability** | FCC | FCC + BCC | BCC (with B2-type SRO) |
| **Thermodynamic linear response prediction** | SRO driven by Co-Cr pair [very weak tetragonal] | SRO driven by Cr-Al pair | SRO driven by Cr -Mn pair |
| **Phase formation (Experiment)** | FCC + tetragonal | FCC + BCC + tetragonal | BCC |
| **Chemical distribution (Experiment)** | Cr-rich phase separation | Cr-rich phase separation | Cr-Mn-Fe and Co-Al based separation |

Table I: Summary of phase stability and SRO predictions by KKR-CPA and experimental observations for the $(CoCrFeMn)_{100-x}Al_x$ HEA system.